\documentclass[a4paper,11pt,russian,ukrainian,english]{article}
\usepackage{iftex}

\iftutex
  \usepackage{babel}
  \babelfont{rm}[Ligatures=Common]{CMU Serif}
  \babelfont{sf}[Ligatures=Common]{CMU Sans Serif}
  \babelfont{tt}{CMU Typewriter Text}
\else
  \usepackage[T2A,T1]{fontenc}
  \usepackage[utf8]{inputenc}
  \usepackage{babel}
\fi

\babeltags{english=english}
\babeltags{russian=russian}

\begin{document}

\begin{center}
\date{}
{THE DEUTERON CHARGE RADIUS $R_C$ IN THE FRAMEWORK OF THE HARD-WALL ADS/QCD MODEL} 

\textrussian{РАДИУС ЗАРЯДА ДЕЙТРОНА $R_C$ В РАМКАХ МОДЕЛИ ЖЕСТКОЙ СТЕНЫ АДС/КХД}

\vskip 1. cm

{Shahin Mamedov $^{a,b}$\footnote{E-mail: sh.mamedov62@gmail.com}}
{Narmin Akbarova $^{a,}$\footnote{E-mail: nerminh236@gmail.com}},
{Minaya Allahverdiyeva $^{b}$\footnote{E-mail: minaallahverdiyeva@ymail.com}}
\vskip 0.5cm


{\textrussian{Шахин Мамедов $^{a,b}$}},
{\textrussian{Нармин Aкбарова $^{a,}$}},
{\textrussian{Минайя Аллахвердиева $^{b}$}}
\vskip 0.5cm

{\it $^a\,$ Institute for Physical Problems, Baku State University, AZ 1148, Z.Khalilov, 23, Baku, Azerbaijan}

{\it $^a\,$\textrussian{Институт физических проблем Бакинского Государственного Университета, Аз 11 48, З.Халилов 23, Баку, Азербайджан}}

{\it  $^b\,$ Institute of Physics, Azerbaijan National Academy of Sciences, AZ 1143, H.Javid av., 131, Baku, Azerbaijan }

{\it  $^b$\textrussian{Институт физики Национальной Академии Наук Азербайджана, Аз 1143, пр. Г.Джавида, 131 Баку, Азербайджан}}

\end{center}

\thispagestyle{empty}

\begin{abstract}
\textrussian{Мы изучаем зарядовый радиус дейтрона в рамках модели жесткой стены AдС/КХД. Мы представляем основные элементы жесткой модели, пишем метрику для пространства AдС. Мы вводим векторное поле с твистом $\tau=6$, описывающее дейтрон внутри пространства AдС, и другое векторное поле для описания фотона соответственно, пишем эффективное действие для взаимодействия объемных полей, находим $G_1(Q^2)$, $G_2(Q^2)$ и $G_3(Q^2)$ формфакторы, затем квадрупольный $G_Q(Q^2)$ и зарядовый $G_C(Q^2)$ формфакторы дейтрона. Таким образом, из зарядового форм-фактора $G_C(Q^2)$ мы находим зарядовый радиус дейтрона $R_C$ в рамках модели жесткой стены AДС/КХД. Затем мы сравниваем наш результат с результатами модели мягкой стены и экспериментальными данными.}
\vspace{0.2cm}

This is English translation of the abstract

We study deuteron charge radius in the framework a hard-wall AdS/QCD model. We present basic elements of the hard-wall model, write metric for the AdS space. We introduce a vector field with twist $\tau=6$ describing deuteron in the bulk of AdS space and other vector field to describe photon respectively, write an effective action for the bulk fields interactions, find a $G_1(Q^2)$, $G_2(Q^2)$ and $G_3(Q^2)$ form factors, then quadrupole $G_Q(Q^2)$ and charge $G_C(Q^2)$ form-factors of a deuteron. Thus, from the charge $G_C(Q^2)$ form-factor we find the deuteron charge radius $R_C$ in the framework of a hard-wall AdS/QCD model. Then we compare our result with the results soft-wall model and experimental data.
\end{abstract}
\vspace*{6pt}


\section*{Introduction}    \label{sec:intro}

Deuterium is one of the two stable isotopes of hydrogen(the other being protium, or hydrogen-1). The nucleus of a deuterium atom, called a deuteron, contains one proton and one neutron, whereas the far more common protium has no neutrons in the nucleus.

Deuteron is often regarded as a loosely bound state of the proton and neutron and so, the study of the deuteron properties can shed light on the structure of the nucleon as well as nuclear effects. The two constituents $-$ proton and neutron inside the deuteron are dominated by the relative S-wave and the D-wave is only about $5$ \%. So, to study deuteron various coupling constants, form factors and charge radius have great significant for nuclear physics.

In this work, we study deuteron charge radius in the framework a hard-wall AdS/QCD model. Notice that, within soft-wall model this parameter is studied only at a zero temperature \cite{Gutsche:2016 ve, Gutsche:2015 T.}. In the soft-wall model confinement and chiral symmetry breaking properties of QCD and finiteness condition of the 5-dimensional (5D) action are provided by introducing an exponential factor called dilaton term, which suppresses expressions under the integral over the extra dimension at infinite values of this dimension (IR) boundary. But within hard wall model this properties are carry out by cutting AdS space at infrared boundary value of fifth coordinate.

In section II, we present basic elements of the hard-wall model, write metric for the AdS space. We introduce a vector field with twist $\tau=6$ describing deuteron in the bulk of AdS space and other vector field to describe photon respectively, write an effective action for the bulk fields interactions, find a $G_1(Q^2)$, $G_2(Q^2)$ and $G_3(Q^2)$ form factors, then quadrupole $G_Q(Q^2)$ and charge $G_C(Q^2)$ form-factors of a deuteron. Thus, from the charge $G_C(Q^2)$ form-factor we find the deuteron charge radius $R_C$ in the framework of a hard-wall AdS/QCD model. Then we compare our result with the results soft-wall model and experimental data.

\label{sec:preparation}
\section*{Paper Preparation}
\section{The hard-wall model of AdS/QCD}

The 5D action in the bulk of AdS space within the hard-wall model is written as\cite{Gutsche:2015 T.}:
\begin{equation} \label{eq 1}
S=\int_{0}\limits ^{z_m} d^4x dz\sqrt{G} \mathcal{L}\left(x,z\right),
\end{equation}
where $G=|\det g_{MN}|$ is the determinant of the $g_{MN}$ metric of the AdS space $(M,N=0,1,2,3,5)$, $\mathcal{L}\left(x,z\right)$ is the Lagrangian and the $z$ coordinate varies in the range $\epsilon\leq z < z_m$, from the $\epsilon\rightarrow0$ ultraviolet (UV) to the IR boundary of AdS space.
The $AdS_{5}$ is written as \cite{Gutsche:2016 ve}:
\begin{equation} \label{eq 2}
ds^{2}=\frac{R^{2}}{z^{2}}\left(-dz^{2}+\eta_{\mu \nu}dx^{\mu}\mu dx^{\nu}\right)=g_{MN}dx^{M}dx^{N},\quad \mu,\nu=0,1,2,3.
\end{equation}
where $\eta_{\mu\nu}=diag(1,-1,-1,-1)$ is a 4D Minkovski metric.

\section{The deuteron charge radius $R_C$}
\bigskip
Here we calculate the deuteron charge radius within the hard-wall model of AdS/QCD. We have used from the general an effective action, which include all interactions between the EM and deuteron fields in the bulk of the AdS space and used in \cite{Gutsche:2016 ve, Gutsche:2015 T.} within the soft-wall model at $T=0$ temperature, and in \cite{Narmin:2022 Huseynova.} within the hard-wall model that contribute to the $G_{i}\left( Q^{2}\right)$, $i=1,2,3$ form factors of a deuteron:
\begin{eqnarray} \label{eq 3}
\textit{S}_{int.}=\int d^{4}x\ \int\limits_{0}^{z_M}dz \sqrt{G}\left[-D^M F_N^+ \left(x,z\right) D_{M} F^{N}\left(x,z\right)-\nonumber \right. \\
-ic_{2}F^{MN}\left(x,z\right) F^{+}_{M}\left(x,z\right)F_N\left(x,z\right)+ \nonumber \\
+\frac{c_{3}}{4M^2_D}e^{2A\left(z\right)}\partial^MF^{NK}\left(x,z\right)\left(i\partial_{K}F^{+}_{M}\left(x,z\right)F_N\left(x,z\right)-\nonumber \right. \\
-F^{+}_{M}\left(x,z\right)i\partial_K F_N\left(x,z\right)+H.C.\left.\left.  \right)  \right].
\end{eqnarray}
In Eq.(3) we perform a Kaluza-Klein decomposition for the vector AdS field with twist $ \tau=6$ dual to the deuteron:
\begin{equation} \label{eq 4} 	F^{\nu}\left(x,z\right)=e^{-\frac{A\left(z\right)}{2}}\sum_{n}F_{n}^{\nu}\left(x\right) \Phi_{n}\left(z\right).
\end{equation}
Then we apply a Fourier transformation for the vector $V\left(x,z\right)$ and deuteron $F_{\nu}\left(x,z\right)$, $F_{\nu}^{+}\left(x,z\right)$ fields respectively and thus, the action terms is written as follows:
\begin{eqnarray} \label{eq 5}
\textit{S}_1=-\left(2\pi\right)^{4}\int\frac{d^{4}p}{\left(2\pi\right)^{4}}\int\frac{d^{4}p^{'}}{\left(2\pi\right)^{4}}\int\frac{d^{4}q}{\left(2\pi\right)^{4}}\delta^{4}\left(p+q-p{'}\right)eV_{\mu}\left(q\right)\int dz V\left(q,z\right)\phi^{2}\left(z\right) \times \nonumber\\
\times\epsilon^{+}(p^{'})\epsilon\left(p\right)(p+p^{'})^{\mu},
\end{eqnarray}
\begin{eqnarray} \label{eq 6}
\textit{S}_2=-c_{2}\left(2\pi\right)^{4}\int\frac{d^{4}p}{\left(2\pi\right)^{4}}\int\frac{d^{4}p^{'}}{\left(2\pi\right)^{4}}\int\frac{d^{4}q}{\left(2\pi\right)^{4}}\delta^{4}(p+q-p{'}) V_{\mu}\left(q\right)\int dz V\left(q,z\right) \phi^{2}\left(z\right)\times  \nonumber\\
\times(\epsilon^{\mu}\left(p\right)\epsilon^{+}(p^{'}) q-\epsilon^{+\mu}(p^{'})\epsilon\left(p\right)\cdot q),
\end{eqnarray}
\begin{eqnarray} \label{eq 7}
\textit{S}_3=\left(2\pi\right)^{4} \frac{c_{3}}{2M^{2}} \int\frac{d^{4}p}{\left(2\pi\right)^{4}}\int\frac{d^{4}p^{'}}{\left(2\pi\right)^{4}}\int\frac{d^{4}q}{\left(2\pi\right)^{4}}\delta^{4}(p+q-p{'}) V_{\mu}\left(q\right)  \int dz V\left(q,z\right) \phi^{2}\left(z\right)\times  \nonumber\\
\times\epsilon^{+}(p^{'})\cdot q\epsilon\left(p\right)\cdot q(p+p^{'})^{\mu}.
\end{eqnarray}
Notice that, the deuteron's EM current is found according to the AdS/CFT correspondence as follows \cite{Gutsche:2016 ve, Gutsche:2015 T.}:
\begin{equation} \label{eq 8}
\left\langle J^{\mu}\left(p,p^{'}\right)\right\rangle=-\frac{\delta e^{iS_{int}}}{\delta V_{\mu}\left(q\right)}|_{V_{\mu}=0}.
\end{equation}
Thus the action terms Eqs.(5)-(7) contribute to the $G_{i}(Q^{2})$, $i=1,2,3$ form factors of the deuteron as follows:
\begin{eqnarray} \label{eq 9}
G_i(Q^2)=c_{i}\int dz V\left(q,z\right) \phi^{2}\left(z\right),
\end{eqnarray}
where $c_{1}=1$, and we fix $c_{2}$ and $c_{3}$ as $c_{2}=1.67066$ and $c_{1}=22.507$ \cite{Narmin:2022 Huseynova.}. Since, the deuteron is a spin-1 particle and current conservation, $P$- and $C$-invariances provide by it, it has three EM form factors which are called the charge $G_C(Q^2)$, quadrupole $G_Q(Q^2)$ and magnetic $G_M(Q^2)$ form factors and the EM current of a $e+d\rightarrow e+d$ electron-deuteron elastic scattering process is written in terms of $G_{i}\left(Q^{2}\right)$ form-factors as follows:
\begin{eqnarray} \label{eq 10}
J^{\mu}(p,p^{'})=-\left(G_{1}\left(Q^{2}\right)\epsilon^{+}(p^{'})\cdot \epsilon\left(p\right)-\frac{G_{3}\left(Q^{2}\right)}{2M^2_{D}}\epsilon^{+}(p^{'})\cdot q  \epsilon\left(p\right)\cdot q \right)(p+p^{'})^{\mu}-\nonumber \\
-G_{2}\left(Q^{2}\right)\left(\epsilon^{\mu}\left(p\right) \epsilon^{+}(p^{'})\cdot q-\epsilon^{+\mu}(p^{'}) \epsilon\left(p\right)\cdot q\right).
\end{eqnarray}

Thus, the charge $G_C(Q^2)$ and quadrupole $G_Q(Q^2)$ form factors of the deuteron in the framework of hard-wall model AdS/QCD are related to the $G_1(Q^2)$, $G_2(Q^2)$ and $G_3(Q^2)$ form-factors as~ \cite{Gutsche:2016 ve, Gutsche:2015 T.}:
\begin{eqnarray} \label{eq 11}
G_C(Q^2)=G_1(Q^2)+\frac{2}{3}\eta_{d}G_Q(Q^2), \nonumber \\
G_Q(Q^2)=G_1(Q^2)-G_2(Q^2)+\left(1+\eta_{d}\right)G_3(Q^2), 
\end{eqnarray}
where $\eta_{d}=\frac{Q^2}{4m^2_D}$, $m_D$ is the deuteron mass.

We derive charge radius of a deuteron from the $G_C(Q^2)$ charge form factor in the framework of the hard-wall model AdS/QCD:
\begin{equation} \label{eq 12}
R_C=\left(-6\frac{dG_C(Q^2)}{dQ^2}|_{Q^2=0}/G_C(0)\right)^{1/2},
\end{equation}
where $G_C(0)=1$.

We calculate these derivatives by use of MATHEMATICA package and then compare obtained results of the values of $R_C$ magnetic radius of a deuteron with the experimental data for this constant and with the soft-wall model results\cite{Gutsche:2016 ve, Gutsche:2015 T.}.
\bigskip
\begin{center}
	{\footnotesize  Table 1.
		
		\bigskip
		\begin{tabular}{|p{0.7in}|p{0.7in}|p{0.7in}|} \hline $R_C^{exp}$(fm) & {$R_C^{s.w.}$ (fm)} &
			$R_C^{h.w.}$ (fm) \\ \hline  2.13$\pm$0.01 & 1.92 & 1.07956 \\
			\hline
		\end{tabular}}
	\end{center}

As is seen from the table our result for the charge radius is less than the result obtained in the soft-wall model framework. This is connected with the fact that the charge form factor obtained in the hard-wall framework is less than one in the soft-wall model \cite{Narmin:2022 Huseynova.}.

\end{document}